\documentclass[aps,prl,twocolumn,superscriptaddress,showpacs]{revtex4}
\usepackage{graphicx}
\begin{document}

\title{Fully three dimensional breather solitons can be created using Feshbach resonance}

\author{M.~Matuszewski}
\affiliation{Institute of Theoretical Physics, Physics Department, Warsaw University,
Ho\.{z}a 69, PL-00-681 Warsaw, Poland}

\author{E.~Infeld}
\affiliation{Soltan Institute for Nuclear Studies, Ho\.{z}a 69,
PL-00-681 Warsaw, Poland}

\author{B.~A.~Malomed}
\affiliation{Department of Interdisciplinary Sciences, School of Electrical
Engineering, Faculty of Engineering, Tel Aviv University, Tel Aviv 69978, Israel}

\author{M.~Trippenbach}
\affiliation{Institute of Theoretical Physics, Physics Department, Warsaw University,
Ho\.{z}a 69, PL-00-681 Warsaw, Poland}

\begin{abstract}
We investigate the stability properties of breather solitons in a three-dimensional
Bose-Einstein Condensate with Feshbach Resonance Management of the scattering length and
confined only by a one dimensional optical lattice. We compare regions of stability in
parameter space obtained from a fully 3D analysis with those from a quasi two-dimensional
treatment. For moderate confinement we discover a new island of stability in the 3D case,
not present in the quasi 2D treatment. Stable solutions from this region have nontrivial
dynamics in the lattice direction, hence they describe fully 3D breather solitons. We
demonstrate these solutions in direct numerical simulations and outline a possible way of
creating robust 3D solitons in experiments in a Bose Einstein Condensate in a
one-dimensional lattice. We point other possible applications.
\end{abstract}
\pacs{03.75.Lm, 05.45.Yv}

\maketitle

The creation of Bose--Einstein condensates (BEC) in vapors of alkali metals has opened an
excellent opportunity to investigate nonlinear interactions of atomic matter waves. An
important challenge of practical interest is to develop methods to create and control
matter-wave solitons. One dimensional dark \cite{1fsol}, bright \cite{2bright}, and
gap-mode \cite{ober} solitons have already been observed. A promising approach to obtain
multidimensional solitons consists in varying the scattering length {\it a} of
interatomic collisions. This can be achieved by means of sweeping an external magnetic
field through the $a=0$ point (the point, where the scattering length vanishes). This
point occurs close to the Feshbach resonance \cite{3Inouye}. The application of an ac
magnetic field may induce a periodic modulation of {\it a}, opening a way to
``Feshbach-resonance management" (FRM) \cite{4Greece}. A noteworthy FRM-induced effect is
the possibility of creating self-trapped oscillating BEC solitons (breathers) without an
external trap in the 2D case.  The underlying mechanism is fast modulations creating an
effective potential on a slower timescale. This potential can stabilize the soliton. The
BEC model based on the Gross-Pitaevskii equation (GPE) with harmonic modulation of {\it
a} was investigated in Refs.~\cite{6Saito,8Abdul,9VPG}. The conclusion was that FRM
renders it possible to stabilize 2D breather solitons even without the use of an external
trap. According to these references, 3D breathers require at least a tight, one
dimensional harmonic trap \cite{Gerlitz}, practically reducing the problem to 2D. Later
on we will call this approach a quasi two dimensional (Q2D) treatment. It will be defined
more precisely below.

Recently \cite{10EL} we demonstrated  that quasi 2D solitons can be stabilized by a
combination of FRM and a strong 1D optical lattice (1D OL), instead of a 1D harmonic trap
\cite{6Saito,8Abdul,9VPG}. By a ``strong lattice" we mean one in which the atoms in
neighboring cells cannot interact. This issue has practical relevance, as a 1D OL can
easily be created, illuminating the BEC by a pair of counterpropagating laser beams that
form a periodic interference pattern \cite{11OL}. The lattice will be
weak or strong depending on the intensity of the laser light. In fact, it is easier to realize a
tight confinement configuration in an optical lattice than in an harmonic trap. Hence
this environment may be more friendly for creating quasi 2D solitons.

In this Letter, we demonstrate that the combined OL-FRM stabilization of 3D solitons is possible
even in a weak lattice, when atoms confined in different cells interact. By analyzing the
stability charts in configuration space we discover two distinct regions where stable solutions
exist. The first of these regions has its counterpart in the Q2D treatment. The other region
appears when the frequency of modulation exceeds a critical value dependent on the
confining potential.
It is not present in the Q2D treatment; hence it corresponds to fully 3D solitons. In the
limit of tight confinement the latter region moves to extremely high frequencies and the
Q2D stability chart is recovered.

\begin{figure}[tbp]
\includegraphics[width=8.5cm]{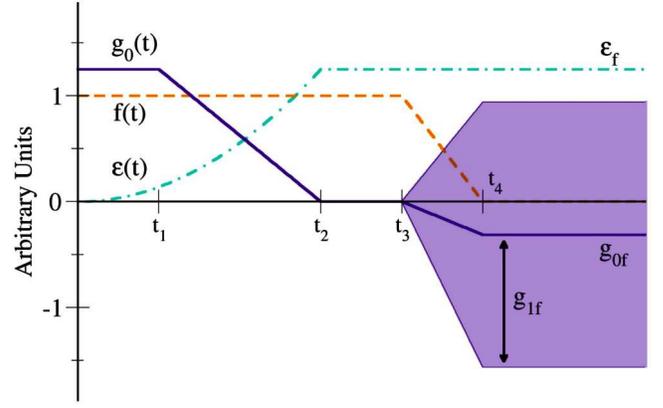}
\caption{(Color online). The time dependence of the nonlinear coefficient, $g$, switching-off function, $f(t)$,
and optical-lattice strength, $\protect\varepsilon ,$ in the numerical experiment which leads to
the establishment of stable 3D breathing solitons supported by the combination of the quasi-1D
lattice and Feshbach-resonance management (FRM). The shaded area indicates rapid oscillations of
$g(t)$, which account for the FRM. \label{fig1}}
\end{figure}

We describe our system by the GPE in physical units, including a
time-dependent (FRM-controlled) scattering length $a(\tau)$ and an
external potential $\tilde{U}(\mathbf{r},\tau)$
\begin{equation}
i \hbar\frac{\partial \Psi}{\partial \tau}=\left[ -\frac{\hbar^2}{2m}\nabla
^{2}+\tilde{U}(\mathbf{r},\tau)+ \frac{4 \pi a(\tau) \hbar^2}{m}|\Psi|^{2}\right] \Psi .
\label{NLS0}
\end{equation}
Initially the BEC is in the ground state of a radial (2D) parabolic trap with frequency
$\tilde{\omega}_{\perp}$, supplemented, in the longitudinal direction, by ``end caps"
induced by transverse light sheets. The configuration is much like the one used to create
soliton trains in a Li${^{7}}$ condensate \cite{2bright}. A 1D lattice potential in the
axial direction is adiabatically turned on from $\tilde{\varepsilon}=0$ to
$\tilde{\varepsilon}=\tilde{\varepsilon}_{\mathrm{f}}$,  see Fig.~\ref{fig1}. Thus, the full
potential is
\begin{eqnarray}
\tilde{U}(\mathbf{r},\tau)&=&\tilde{\varepsilon} (\tau)\left[ 1-\cos (2 \pi z/\lambda)\right] + \nonumber \\
&+&f(\tau)\left[(m/2)\tilde{\omega}_{\perp}^{2}\varrho ^{2}+\tilde{U}_{0}(z)\right],
\label{V0}
\end{eqnarray}
where $\lambda$ is the lattice spacing, $\varrho$ is the radial variable in the plane
transverse to $z$, and the axial ``end-cap'' potential, $\tilde{U}_{0}(z)$ is
approximated by a sufficiently deep one dimensional rectangular potential well. The width
of the well determines the number of peaks in the finally established structure. The
$f(\tau)$ is a switching-off function (see Fig.~\ref{fig1}). We introduce dimensionless
variables $\mathbf{x}=(\pi/\lambda)\mathbf{r}$, $t=\tau \omega$, $\psi=\Psi
\sqrt{\lambda^3/(N\pi^3)}$, where $\omega=\pi^2 \hbar/(m \lambda^2)$ and $N$ is the
number of atoms. We also define $g=4 \pi^2 N a /\lambda$,
$\varepsilon=\tilde{\varepsilon}/(\omega\hbar)$, $\omega_{\perp}=\tilde{\omega}_{\perp
}/\omega$, and $U_0=\tilde{U}_{0}/(\omega\hbar)$ to obtain
\begin{equation}
i\frac{\partial \psi}{\partial t}=\left[ -\frac{\nabla ^{2}}{2}+U(\mathbf{x},t)+g(t)|\psi
|^{2}\right] \psi , \label{NLS}
\end{equation}
and the potential in the form
\begin{equation}
U(\mathbf{x},t)=\varepsilon (t)\left[ 1-\cos (2z)\right] +f(t)\left[\frac{1}{2}\omega _{\perp
}^{2}\varrho ^{2}+U_{0}(z)\right].  \label{V}
\end{equation}
The nonlinear interaction coupling is described by
\begin{equation}
g(t)=g_{0}(t)+g_{1}(t)\sin (\Omega t),  \label{g}
\end{equation}
and the dimensionless modulation frequency by
$\Omega=\tilde{\Omega}/\omega$. Initially $g_1(0)=0$ and $g(0)=g_0(0)>0$.
At some moment $t_{1}$, we begin to linearly decrease $g_0(t)$. It
vanishes at time $t_{2}$, and remains zero up to $t_{3}$, when we
start to gradually switch on the rapid FRM modulation of {\it a}. In
the interval $[t_3,t_4]$, $g_0(t)$ decreases linearly from zero to a
negative $g_{0 \mathrm{f}}$ and the amplitude of the modulation
$g_1(t)$ increases from zero to $g_{1 \mathrm{f}}$, see
Fig.~\ref{fig1}. Simultaneously,  both the radial confinement and
end-caps are gradually switched off.
At times $t>t_{4}$, $g(t)$ oscillates with a constant amplitude $g_{1 \mathrm{f}}$ around a
negative average value $g_{0 \mathrm{f}}$. Consequently, a soliton so created, if any,
is supported by the combination of the 1D lattice and FRM.

\begin{figure}[tbp]
\includegraphics[width=8.5cm]{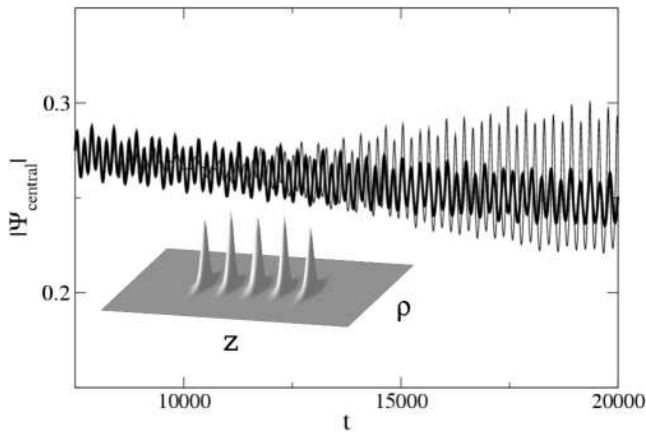}
\caption{Evolution of the amplitude of the central peak in the pattern comprising five cells of
the optical lattice (shown in the inset). The normalized parameters are $g_{0 \mathrm{f}}=-30$,
$g_{1 \mathrm{f}}=4g_{0 \mathrm{f}}$, $\varepsilon_{\mathrm{f}}=20.5$, $\Omega =22$, $\protect\omega _{\perp }=0.3$,
$t_{1}=30$, $t_{2}=100$, $t_{3}=120$, and $t_{4}=130$. The snapshots in the inset is taken at
$t=6000$. The unit of time is $m\lambda^2/(\pi^2\hbar$). \label{fig2}}
\end{figure}

Numerical experiments following the path outlined in Fig.~\ref{fig1} indicate that it is possible
to create stable solitons \cite{10EL} (see inset to Fig.\ref{fig2}). Before showing the
results, we first resort to the variational approximation (VA) in order to predict conditions on
the modulation frequency and the size of the negative average nonlinear coefficient $g_{0 \mathrm{f}}$,
necessary to support 3D solitons.


The VA can be applied to the description of BEC dynamics under diverse circumstances
\cite{6Saito,8Abdul,9VPG,10EL,12BBB,Progress}. Equation (\ref{NLS}) is derived from the Lagrangian
density
\begin{eqnarray}
{\cal{L}}= i(\psi _{t}^{\ast }\psi -\psi _{t}^{\ast }\psi )-|\psi _{\varrho }|^{2}-|\psi _{z}|^{2}
-g(t)|\psi |^{4} - 2U|\psi |^{2} \label{L}
\end{eqnarray}
We use VA for $t>t_3$ and choose a complex Gaussian ansatz for the solution for one
lattice cell. The amplitude is $A(t)$, radial and axial widths are $W(t)\ $ and $V(t)$
respectively, and $b(t)$ and $\beta (t)$ are the corresponding chirps
\begin{equation}
\psi (\mathbf{r},t)=Ae^{\left[ -\varrho^{2}(1/2W^{2}+ib) -z^{2}( 1/2V^{2}+i\beta)+i\phi\right]}.
\label{ansatz}
\end{equation}
The reduced Lagrangian can be found upon substituting (\ref{ansatz}) into (\ref{L}) and
integrating over space.
By varying this reduced Lagrangian with respect to $\phi$ we obtain the constant
$E=A^{2}W^{2}V =\pi ^{-3/2}\int_{cell}|\psi |^{2}d\mathbf{r}=\pi ^{-3/2}/n$, where the
integral extends over one cell of the lattice and $n$ is the number of occupied lattice
cells. Notice that the total number of atoms is included in the definition of the
nonlinear coupling $g(t)$ and the total wavefunction is normalized to unity. When the
other four variational equations are derived, we can deduce two dynamical equations for
the widths:
\begin{eqnarray}
\ddot{W} &=&\frac{1}{W^{3}}-f(t)\omega _{\perp }^{2} W+ \frac{Eg(t)}{\sqrt{8}W^{3}V},
\label{variat1} \\
\ddot{V} &=&\frac{1}{V^{3}}-4\varepsilon_{\mathrm{f}} V\exp \left( -V^{2}\right)
+\frac{Eg(t)}{\sqrt{8}W^{2}V^{2}}.  \label{variat2}
\end{eqnarray}
These equations describe the dynamics of a single peak, and with one term slightly altered can
be applied to the problem of a BEC confined in a 1D harmonic trap \cite{6Saito,8Abdul,9VPG}.

In the corresponding Q2D treatment we drop the $z$ dimension in Eq.~(\ref{NLS}). It is
assumed that in this direction the profile of the wavefunction is fixed and reproduces
the ground state $\psi_0$ of the single lattice cell (Wannier function) or harmonic
potential, as in Refs.~\cite{6Saito,9VPG}. The reduced potential in 2D GP will take the
form $U(\varrho,t)= f(t)(1/2)\omega _{\perp }^{2}\varrho ^{2}$. In VA we take $V \equiv
V_0$ from $4\varepsilon_{\mathrm{f}} V_{0}^{4}\exp \left( -V_{0}^{2}\right)=1$, and only
solve equation (\ref{variat1}). In numerical simulations we rescale the nonlinear
coupling coefficient $g_{\mathrm{2D}} = g \times (\int |\psi_0|^2 \psi_0 dz)/(\int \psi_0
dz)$.


We simulated both the full GPE, Eq.~(\ref{NLS}), using an axisymmetric code (for 3D), a
Cartesian code (for 2D), and the variational equations for comparison. Numerical
simulations followed the path outlined in Fig.~\ref{fig1}. The parameters used in the
simulations would correspond, for $^{85}$Rb atoms and an OL period of $\lambda = 3\,\mu
$m, to an initial radial-confinement frequency of $\omega _{\perp }=2 \pi \times 39$ Hz,
a FRM frequency of $\Omega=2 \pi \times 2.87$ kHz. A good candidate is the Feshbach
resonance at 155 G for $^{85}$Rb. Since in this case the $a=0$ point is far from the
resonance, both two and three body losses are negligible \cite{155G}. A lattice depth of
$\varepsilon_{\mathrm{f}} =10.25$ recoil energies, and an effective nonlinear coefficient
of $(N/n)a=\pm \, 10^{-5}\,$m are required. Here $N/n$ is the number of atoms per lattice
cell, with a {\it total} number of atoms in the range of $10^{4}-10^{6}$. The respective
values of the normalized parameters are given in the figure captions.

Examples of the numerical results, which are generic, are displayed in Fig.~\ref{fig2},
which shows the evolution of the central-peak's amplitude versus time in dimensionless
units, defined by Eq.~(\ref{NLS}). After an initial transient, a stable structure is
established, featuring breathing without any systematic decay. The figure reveals the
influence of neighboring solitons on the amplitude of the central peak. The thin curve
corresponds to the evolution of the full multipeak structure, shown in the inset for a
fixed moment of time. To obtain the thick curve we repeated the above calculations up to
the time $t=7500$ and then removed all but the central peak. In this case the amplitude
gradually decreased. The interaction between neighboring solitons can be explained by the
difference between oscillation periods from cell to cell due to small deviations in the
numbers of atoms. It is essentially the same mechanism as in a Josephson junction. When
the lattice potential is weak, interaction between neighboring solitons is possible.

\begin{figure}[tbp]
\includegraphics[width=8.5cm]{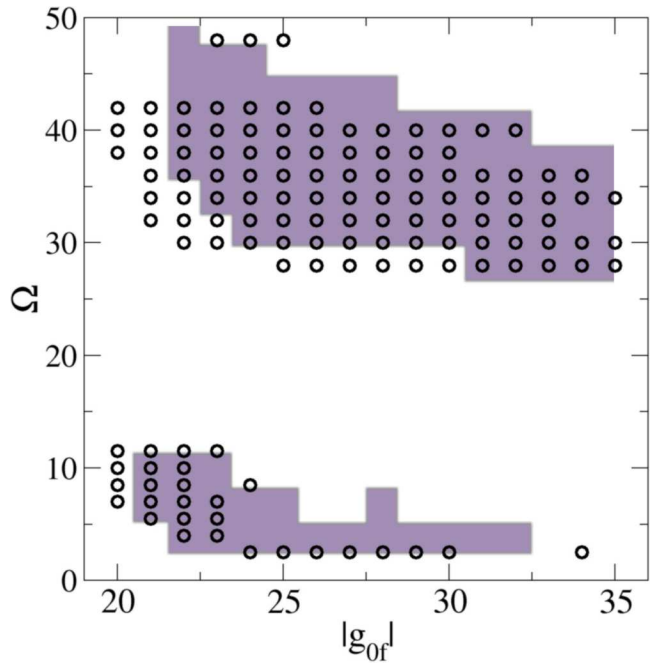}
\caption{(Color online). Stability regions for the 3D solitons in the $\left( |g_{0 \mathrm{f}}|,\Omega \right)$
plane, as predicted by the variational approximation (shaded area),
and found from direct simulations of the Gross-Pitaevskii equation (circles).
Other parameters are as in Fig.~\protect\ref{fig2}, except $\varepsilon_{\mathrm{f}}=50$.
Note the similarity of the lower region to that of Fig.~\protect\ref{fig3b}. This region
corresponds to Q2D solitons. On the other hand, the upper region contains fully 3D
solitons. The borders of the VA stability regions were found analytically and will be
given in a fuller version \cite{future}. The lowest excitation frequency of the confining
potential is $\Omega_0=26.76$.\label{fig3a}}
\end{figure}

\begin{figure}[tbp]
\includegraphics[width=8.5cm]{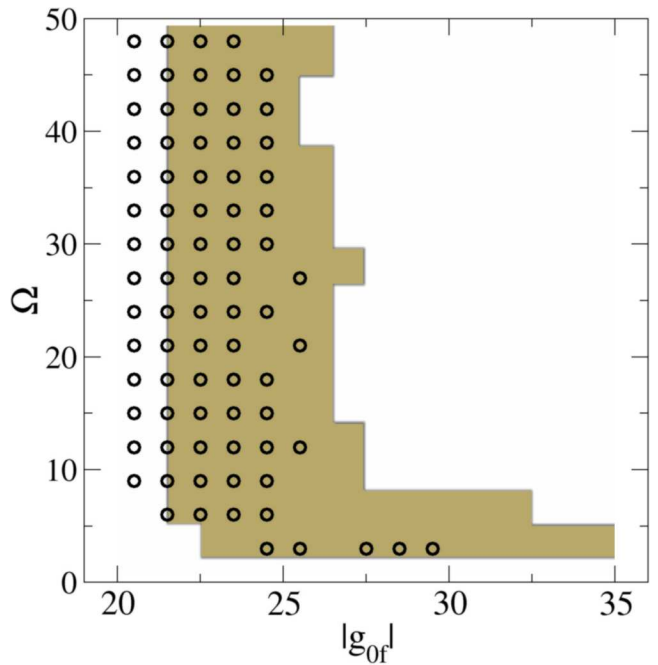}
\caption{(Color online). Same as Fig.~\protect\ref{fig3a}, but for Q2D treatment. \label{fig3b}}
\end{figure}

In Figs.~\ref{fig3a} and \ref{fig3b} we have collected results of a systematic scan of
parameter space based on GPE simulations and compared them with the predictions of the VA
(a similar analysis can be performed if we replace 1D OL with a 1D harmonic trap - the
conclusion does not depend on the form of confinement in the $z$ direction). By stability
we mean shape preservation during one run. When this was not the case, after a short
period of time we observed clear collapse or spreadout in transverse directions.

The agreement between VA and direct simulations is very good. As seen from
Fig.~\ref{fig3a}, in our fully 3D treatment we found two islands of stability. Note the
similarity of the lower region to that of Fig.~\protect\ref{fig3b}, which portrays the
results of the Q2D treatment. This region corresponds to Q2D solitons. On the other hand,
the upper region contains fully 3D solitons. It appears when the frequency of modulation
exceeds the lowest excitation frequency of the confining potential, $\Omega_0=26.76$. If
the strength of the lattice $\varepsilon_{\mathrm{f}}$ is increased, this region moves
towards higher frequencies, the Q2D region expands upwards, and becomes more and more
like in Fig.~\protect\ref{fig3b}. This will be demonstrated in a fuller version of this
work \cite{future}.

As we saw, new region of stability appears in the 3D treatment as compared to the Q2D
treatment. We are more used to the effect of adding a new dimension simply shrinking or
abolishing the basin of stability of solitons or waves. For example, water waves are
unstable with respect to perturbations along their direction of propagation only when the
depth exceeds a critical value \cite{TB}. When, however, two dimensional perturbations
are allowed, there will always be an unstable angle regardless of the depth \cite{WD}.
Another example is that of 1D solitons of the Nonlinear Schr\"{o}dinger equation (NLS)
with constant coefficients. They are stable in 1D, but unstable in 2D or 3D. This is also
true for some NLS waves \cite{EI}.

In the problem treated in this Letter this is not the case. For much of the quasi two
dimensional stability chart adding a degree of freedom {\it stabilizes} the soliton
solution. The key to this dichotomy seems to be the presence of a periodic modulation,
absent in the above mentioned classical examples. This can be illustrated by a simple
case involving oscillators. Take as the one dimensional version a forced oscillator
problem:
\begin{equation}
\ddot{x}+\omega_0^2x=y\cos(\omega_0t).
\end{equation}
If $y$ is fixed, the solution has a secular component $x=yt/(2\omega_0)\sin(\omega_0t) +
F(t)$, where $F(t)$ is a periodic function, and so the amplitude will grow as $t$. If
however we allow a second degree of freedom, such that $y$ also oscillates (for instance
$\ddot{y}+\epsilon^2y=0$) the solution stabilizes, unless $\epsilon = \pm 2\omega_0$. In
general this can be the case when there are periodic modulations. This fact, obvious in
oscillator theory, is perhaps less well known in the soliton context.


The main result of this letter is the possibility of creating fully 3D breather solitons
in a BEC confined by a 1D optical lattice potential, corresponding to the upper region in
Fig.~\ref{fig3a}. The stable patterns may feature a multi-cell structure, which in the
case studied here forms a set of weakly interacting fundamental solitons. The scheme
proposed in this Letter is based on a combination of FRM and a 1D optical lattice, and
could be implemented in an experiment, as outlined here. This would open the way to the
creation of robust 3D solitons (breathers) in BECs. A similar idea can apply in the field
of nonlinear optics. The Feshbach resonance could be replaced by a nonlinear periodic
structure, for example ferromagnetic domains as used in contemporary quasi phase matching
\cite{Arie}, or by optically induced photorefractive lattices \cite{Wiesio}.

M.M. acknowledges support from the KBN grant 2P03 B4325, M.T. was supported by the Polish Ministry
of Scientific Research and Information Technology under grant PBZ MIN-008/P03/2003 and E.I. from
that of grant 2P03B09722. The work of B.A.M. was partially supported by the Israel Science
Foundation through grant No. 8006/03.

\end{document}